\documentclass[3p,times,twocolumn]{elsarticle}

\usepackage{ecrc}

\volume{00}

\firstpage{1}

\journalname{Nuclear Physics B Proceedings Supplement}

\runauth{G. Vujanovic et al.}

\jid{nppp}

\jnltitlelogo{Nuclear Physics B Proceedings Supplement}

\usepackage{amssymb}
\usepackage[figuresright]{rotating}

\begin{document}

\begin{frontmatter}



\dochead{}

\title{Probing the dissipative properties of a strongly interacting medium with dileptons}


\author[McGill]{Gojko Vujanovic}
\author[McGill]{Chun Shen} 
\author[McGill,BNL]{Gabriel S. Denicol}
\author[BNL]{Bj\"orn Schenke}
\author[McGill]{Sangyong Jeon}
\author[McGill]{and Charles Gale}

\address[McGill]{Department of Physics, McGill University, 3600 rue University, Montr\'eal, Qu\'ebec H3A 2T8, Canada}
\address[BNL]{Physics Department, Brookhaven National Lab, Building 510A, Upton, NY, 11973, USA }

\begin{abstract}
We investigate the effects of the presence of a non-vanishing net baryon number density and its diffusion on dilepton production, within a hydrodynamical description of the medium created at $\sqrt{s_{NN}}=7.7$ GeV collision energy. This energy value is explored within the Beam Energy Scan (BES) program at Relativistic Heavy Ion Collider (RHIC) at Brookhaven National Laboratory. Particular attention is devoted to a new dissipative degree of freedom: the net baryon number diffusion ($V^\mu$), and to the net baryon number conductivity ($\kappa$) | a transport coefficients governing the overall magnitude of $V^\mu$. The effects of $\kappa$ on dilepton production are assessed, with an outlook on how future experimental dilepton data can be used to learn more about $\kappa$.
\end{abstract}


\end{frontmatter}

% \linenumbers

\section{Introduction}\label{sec:intro}
%%%%%%%%%%%%%%%%%%%%%%%%%%%%%%%%%%%%%%%%%%%%%%%%%%%%%%%%%%%%%%%%%%%%%%%%%%%%%%%%%%%%%%%%%%%%%%%%%%%%%%%%%%%%%%%%%%%%%%%%%%%%%%%%%%%%%%%%%%%%%%%
The hydrodynamic modeling of the medium created in the BES program at RHIC has been revisited in recent years. Indeed, developments in lattice QCD calculations at finite net baryon chemical potential\footnote{The equation of state used in this study is provided in Ref. \cite{Aki_priv}.}, using a Taylor expansion around $\mu_B=0$ \cite{Borsanyi:2011sw, Bazavov:2012jq}, and in our understanding of relativistic dissipative fluid dynamics are expected to considerably improve our theoretical description of low energy heavy ion collisions.

In fact, given the large ongoing effort to describe the medium from the BES using hydrodynamics, the goal of this proceedings is to study effects observed in dilepton yield and anisotropic flow that stem from the presence of net baryon number density (and its diffusion) within a hydrodynamical evolution. Thus, this is an exploratory study of the influence net baryon chemical potential ($\mu_B$) and baryon diffusion have on dilepton production, with a more quantitative study being planned in the near future as the hydrodynamical description of the medium is perfected.     

%%%%%%%%%%%%%%%%%%%%%%%%%%%%%%%%%%%%%%%%%%%%%%%%%%%%%%%%%%%%%%%%%%%%%%%%%%%%%%%%%%%%%%%%%%%%%%%%%%%%%%%%%%%%%%%%%%%%%%%%%%%%%%%%%%%%%%%%%%%%%%%
\section{Initial conditions and hydrodynamical evolution}\label{sec:IC_hydro}
%%%%%%%%%%%%%%%%%%%%%%%%%%%%%%%%%%%%%%%%%%%%%%%%%%%%%%%%%%%%%%%%%%%%%%%%%%%%%%%%%%%%%%%%%%%%%%%%%%%%%%%%%%%%%%%%%%%%%%%%%%%%%%%%%%%%%%%%%%%%%%%
These proceedings focus on heavy-ion collisions at $\sqrt{s_{NN}}=7.7$ GeV. The initial energy density in the transverse direction is distributed according to the average distribution obtained from 1000 MC-Glauber events. The events are averaged in such a way that the event plane angle $\Psi_2$ is always aligned with the x-axis. This procedure guarantees a more realistic initial state eccentricity, leading to a better description of the final state elliptic flow. The net baryon density profile in the transverse plane has the same shape as the energy density, while its normalization is tuned such that the number of participants $N_{\rm part}$ obtained from the Glauber model is reproduced for $\sqrt{s_{NN}}=7.7$ GeV and 0-80\% centrality class. In the longitudinal direction, the normalized net baryon density profile is given by:
\begin{eqnarray}
g_B &=& N \left\{ \Theta(|\eta_s|-\eta_{s,0}) \exp\left[-\frac{(|\eta_s|-\eta_{s,0})^2}{2\Delta\eta^2_{s,1}}\right]\right.\nonumber\\
    &+& \left. \left[1-\Theta(|\eta_s|-\eta_{s,0})\right]\times\right.\nonumber\\
		&\times& \left. \left[A+\left(1-A\right) \exp\left[-\frac{(|\eta_s|-\eta_{s,0})^2}{2\Delta\eta^2_{s,2}}\right] \right] \right\}\nonumber,
\label{eq:init_cond_nB}
\end{eqnarray}
where $\eta_s$ is the usual space-time rapidity and $N$ is chosen such that $\int d\eta_s g_B = 1$. The free parameters $\eta_{s,0}=2.09$, $A=0.8$, $\Delta\eta_{s,1}=0.7$, $\Delta\eta_{s,2}=1$ were chosen so that the net proton rapidity distribution $\frac{dN}{dy}$ is reproduced at $\sqrt{s_{NN}}=7.7$ GeV and 0-80\% centrality. Note that the same functional form is used to initialize the energy density profile in the longitudinal direction, with the overall normalization for energy density fitted to the experimental charged hadron $\frac{dN^{ch}}{d\eta}$ spectrum. Lastly, the initial velocity in $\tau$--$\eta_s$ coordinates is chosen to be $u^\mu=(1,0,0,0)$.

The equation of motion for net baryon number diffusion is:
\begin{eqnarray}
\tau_V \Delta^\mu_\nu u^\sigma \partial_\sigma V^\nu + V^\mu = \kappa \nabla^\mu \alpha_0-\delta_{VV}V^\mu\theta - \lambda_{VV} \sigma^{\mu\nu} V_{\nu}\nonumber
%\label{eq:V_mu}
\end{eqnarray}
where $\alpha_0=\frac{\mu_B}{T}$ and $\theta=\partial_\mu u^\mu$. The transport coefficients are given by $\tau_V=\frac{0.2}{T}$, $\kappa=0.2\frac{n_B}{\mu_B}$, $\delta_{VV}=\tau_V$, $\lambda_{VV}=\frac{3}{5}\tau_V$. $\tau_V$ and $\kappa$ come from the AdS/CFT calculation in Ref. \cite{Natsuume:2007ty}, while the terms $\lambda_{VV}$ and $\delta_{VV}$ are derived using the Boltzmann equation in the 14-moment approximation \cite{Denicol:2012cn}. The initial condition for the net baryon number diffusion is $V^\mu=0$. 

The equation of motion for the shear stress tensor and the transport coefficients that enter this equation are the same as used in Ref. \cite{Vujanovic:2013jpa}. The only exception is the shear viscosity coefficient, where instead of $\frac{\eta}{s}=0.08$, we now employ $\frac{\eta T}{(\varepsilon+P)}=0.08$. Note that in the limit of vanishing net baryon number density, these two prescriptions are the same. The initial conditions $\pi^{\mu\nu}=0$ still hold while the initialization (or thermalization) time is $\tau_0=0.6$ fm/c. The freeze-out energy density of 0.1 GeV/${\rm fm^3}$ was chosen instead of a freeze-out temperature, as it provided a better fit to hadronic observables. 

%%%%%%%%%%%%%%%%%%%%%%%%%%%%%%%%%%%%%%%%%%%%%%%%%%%%%%%%%%%%%%%%%%%%%%%%%%%%%%%%%%%%%%%%%%%%%%%%%%%%%%%%%%%%%%%%%%%%%%%%%%%%%%%%%%%%%%%%%%%%%%%
\section{Results and conclusion}\label{sec:results}
\begin{figure}[!h]
\includegraphics[scale=0.5]{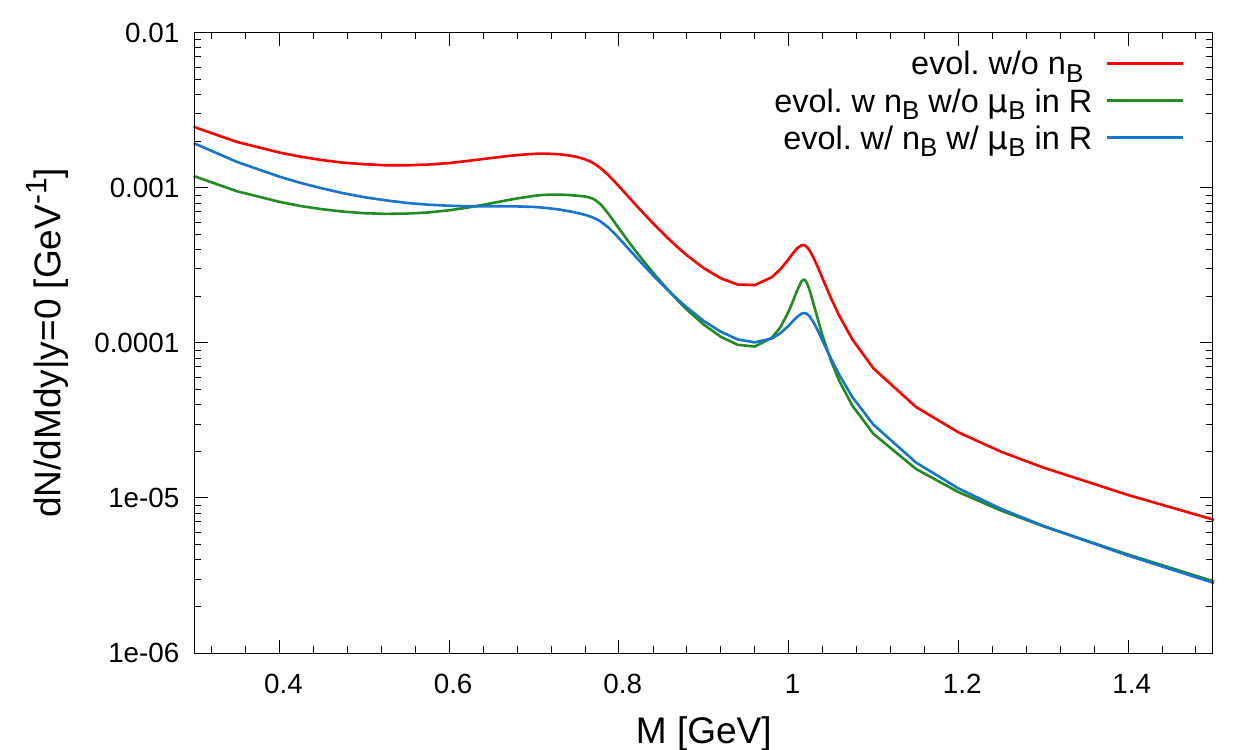}
\includegraphics[scale=0.5]{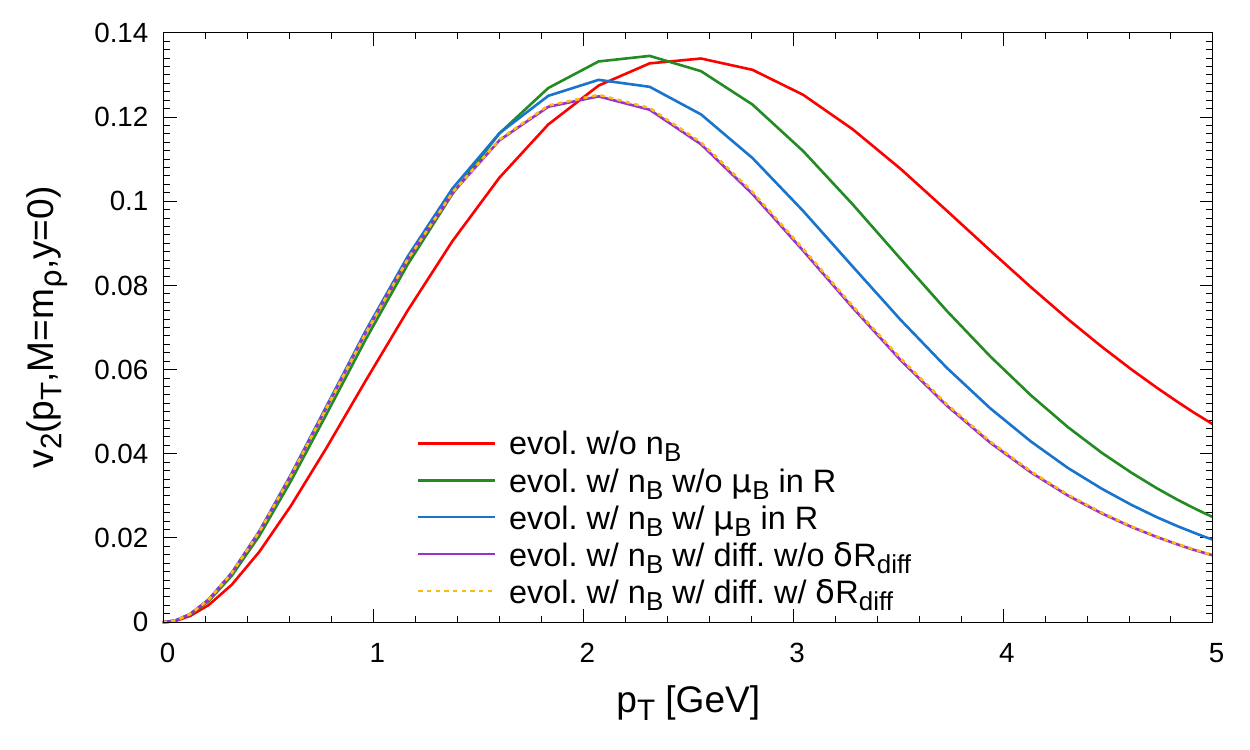}
\caption{Dilepton invariant mass yield (a) and elliptic flow $v_2$  as a function of $p_T$ at fixed $M=m_\rho$ (b) including baryon diffusion effects in the hydrodynamical evolution of the medium. There results assume 0-80\% centrality and $\sqrt{s_{NN}}=7.7$ GeV collision energy.}
\label{fig:dilep_yield_v2_BES_w_diff}
\end{figure}

%%%%%%%%%%%%%%%%%%%%%%%%%%%%%%%%%%%%%%%%%%%%%%%%%%%%%%%%%%%%%%%%%%%%%%%%%%%%%%%%%%%%%%%%%%%%%%%%%%%%%%%%%%%%%%%%%%%%%%%%%%%%%%%%%%%%%%%%%%%%%%%
The dilepton yield and elliptic flow are presented in Fig.~\ref{fig:dilep_yield_v2_BES_w_diff}. In order to obtain the dilepton yield and elliptic flow, the correction $\delta R$ to the thermal production rate accounting for baryon diffusion was computed using a similar procedure as for the viscous correction shown in Ref. \cite{Vujanovic:2013jpa}.  
All the curves shown in Fig.~\ref{fig:dilep_yield_v2_BES_w_diff} start from the same initial energy density profile. The red curve corresponds to a medium not influenced by baryon number, while the green curve corresponds to a medium solely affected by net baryon number density (and no diffusion). As no $\mu_B$ is present in the dilepton rate, the red and green curves in Fig.~\ref{fig:dilep_yield_v2_BES_w_diff} (a) show the sensitivity of dileptons to temperature profile differences induced by the net baryon number degree of freedom. Including $\mu_B$ in the dilepton rates, increases the width of in-medium vector mesons, clearly depicted by the blue curve in Fig.~\ref{fig:dilep_yield_v2_BES_w_diff} (a). The invariant mass dilepton yield isn't significantly affected by diffusion of net baryon number, since the latter doesn't significantly increase $\mu_B$ nor the amount of entropy generation, given the current assumption of the baryon diffusion constant $\kappa$. Hence, $\frac{dN}{dMdy}$ of thermal dileptons isn't affected by baryon diffusion. On the other hand, the $p_T$--differential dilepton elliptic flow in Fig.~\ref{fig:dilep_yield_v2_BES_w_diff} (b) is affected by the presence of both $\mu_B$ and baryon diffusion (see purple/yellow curve relative to the blue curve). This sensitivity can be used in the future to possibly constrain $\kappa$ using dilepton elliptic flow data.

In summary, we have presented a first calculation of dilepton production from a medium at finite $\mu_B$ and $V^\mu$. The thermal dilepton $v_2(p_T)$ results presented are promising as they open the possibility to constrain a new transport coefficient $\kappa$ governing the size of baryon diffusion in strongly interacting media. More in-depth studies on the sensitivity of thermal dileptons to $\kappa$ are in progress. 



%% References

%\nocite{*}
\vspace{-0.195cm}
\bibliographystyle{elsarticle-num}
\bibliography{references}






% \end{thebibliography}

\end{document}